# An Investigation of the Incidence and Effect of Spreadsheet Errors Caused by the Hard Coding of Input Data Values into Formulas

Paul J. Blayney
Faculty of Economics + Business, University of Sydney
p.blayney@econ.usyd.edu.au

**ABSTRACT**

The hard coding of input data or constants into spreadsheet formulas is widely recognised as poor spreadsheet model design. However, the importance of avoiding such practice appears to be underestimated perhaps in light of the lack of quantitative error at the time of occurrence and the recognition that this design defect may never result in a bottom-line error. The paper examines both the academic and practitioner view of such hard coding design flaws. The practitioner or industry viewpoint is gained indirectly through a review of commercial spreadsheet auditing software. The development of an automated (electronic) means for detecting such hard coding is described together with a discussion of some results obtained through analysis of a number of student and practitioner spreadsheet models.

## 1. INTRODUCTION

This paper presents one aspect of a work-in-process PhD proposal that currently has the overarching objective of providing insight into the incidence and effect of spreadsheet errors resulting from deficient spreadsheet design. While this future, doctoral research will examine a number of design issues, particularly the effect of inadequate documentation, the scope of this piece of writing is not nearly as expansive. The focus of this paper is to investigate the incidence and effect of the hard coding of data input values (constants) into formulas. While this practice is widely recognised as a spreadsheet model design fault it appears not to have been extensively studied nor considered to be the source of significant errors.

The importance of this research is predicated by the widespread use of spreadsheets for both generation of financial accounting results and for decision making. Conclusions by authors such as (Croll 2005) that whole industries are run on a spreadsheet and that entire markets are exposed to unreasonable levels of risk due to uncontrolled spreadsheet use are arguably overstated. However the high incidence of spreadsheet errors is well documented and widely reported. For example, Panko and Ordway (2005) present the results of a number of surveys depicting the widespread usage of spreadsheets for critical financial reporting together with the prevalence of spreadsheet errors. The difficulties involved in the identification of spreadsheet errors (Panko, 2000) are also well established.

The hard coding of input data into formulas is the type of error that initially does not result in a quantitative error, e.g. a bottom-line misstatement of a P&L or income statement. And it may never result in a bottom-line error. Furthermore such hard coding will save time in the spreadsheet model development process and result in a slightly less complex model. The question then is whether there really is a significant problem with this type of so-called error.

The first objective of this paper is to determine the perspective currently held in academia and industry with regards to the importance of the hard-coding of input data values in spreadsheet model formulas. The second goal of the current research is to describe the development of an electronic means (i.e. an Excel VBA procedure) for revealing instances of formulas containing hard coded constants or input values. The paper concludes with a discussion of some results obtained from use of this procedure on a number of student and practitioner spreadsheet models.

This study hypothesises that many spreadsheet models contain hard coded data input values that result in significant bottom-line errors the incidence of which being severely underestimated due to the difficulty of detection. It predicts that future use of the developed detection procedure will reveal instances of hard coding that cause such significant bottom-line errors and that their frequency and magnitude will be positively correlated with the length of time that the model has been in use (e.g. how long since its initial development).

**1.1 Background and Motivation**

The impetus for this paper originates from the author's work with accounting students' use of interactive self-assessable spreadsheet assignments prepared using the method described in Blayney and Freeman (2004). These assignments require students to use general cell-referenced formula for all entries in order to have their work assessed as being correct. Hard coding of input data into formulas will result in an assessment that the entry is incorrect. A not infrequent complaint received from students is for their work to be assessed as being wrong even though the result of their entry formula produces the correct result with the data provided.

This occurs when a student has entered a formula for their answer that is not generalisable. The self-assessment marking program determines whether a formula is correct by testing its result with a different data set than the one provided to the student. If the student's formula produces the result expected with the changed data set the entry is evaluated as being correct. If an entry contains hard-coded values instead of the correct cell-referenced cells it will be assessed as being incorrect.

Such hard coding errors are similar in many ways to omission errors with their danger potential and detection difficulty (Panko 2005). The likelihood that hard coded data input values will result in bottom-line errors will increase over time as the developer forgets which items are hard coded or when the model is used by someone other than the developer (especially in the typical case where the hard coding is not documented). The passage of time will also increase the probability that the assumption that has been hard coded into formulas (e.g. VAT rate of 7.5%) will change.

**2. ALTERNATIVE VIEWPOINTS OF HARD CODING**

Insight into the view held by academics towards hard coding will be gained by several means. First of all, current recommended best practice for spreadsheet model development will be overviewed. Included in this review will be a comparison of spreadsheet modelling with traditional program design. An examination of recommended software and spreadsheet engineering methods will be conducted in conjunction with this review. Secondly a review of the academic spreadsheet error research and the error classifications established and used by these researchers will be undertaken.

For this paper the industry viewpoint will be proxied by the position with regards to hard coding taken by commercial spreadsheet auditing software vendors. Future research will gather direct evidence from practitioner interviews and analysis of practitioner spreadsheet models. However, this study will make the assumption that the attitude taken by commercial spreadsheet auditing software vendors will provide a reasonably accurate portrayal of the practitioner or industry attitude towards the hard coding of input data. This assumption is made on the basis that commercial vendors have a strong financial interest to determine the preferences of their customers (industry) and to provide the desired software features.

## 2.1 Comparison to Traditional Programming Best Practice

The problem of hard coding of values into spreadsheet formulae has an equivalent in traditional programming where a well constructed program would rarely (never) enter a value (a number) into a line of executable code. The recommended programming technique is to assign the desired value to a named variable (e.g. RedFont = 3) and use this variable in the program code. This can be contrasted to the inferior design that would simply type the value 3 into a line of code. The prevalence of such inferior design practices are caused by the practice being quicker in the short run as the developer avoids having to create a variable and assign a value to it. The resulting 'magic number' in the code "makes programs harder to read, understand, and maintain" (Wikipedia 2005).

Whereas 'magic numbers' in programming statements are bad programming practice the issue with spreadsheet models is compounded by the variety of ad hoc methods that modellers use in the design and implementation of their models. As with traditional programming, hard coding into spreadsheet formulas also has the serious potential consequence of output errors. Rajalingham's discussion of actual errors as contrasted with potential errors is worthy of note in this paper as the hard-coding errors being considered may or may not result in output errors in the future and will not result in errors at the time of the hard coding. The hard-coding identification method will therefore be useful both for output error detection and as a prevention device for potential future bottom-line misstatements.

This research is concerned with cases where the developer knowingly and deliberately hard-codes values into formulas. This may be done based on the developer's reasoning that the hard-coded value is data that is not subject to change and the recording of the amount as a data item on the model's input sheet is superfluous. Alternately the hard coding may occur as a short term time saving measure with the developer recognising that the value will need to be updated at a later time if circumstances change.

An example of such erroneous reasoning situation might be where a developer hard-codes the number 0.075 for the VAT rate of 7.5% with the reasoning that this amount is not subject to change. If this tax rate is subsequently altered, the hard-coding of the value will result in quantitative errors unless all formulas containing the hard coded 0.075 are updated. The output errors occurring as a result of this design flaw could be classified as domain knowledge errors according to the Rajalingham error classification method.

## 2.2 Spreadsheet Model Development Best Practice

Spreadsheet model development is similar to traditional programming in more than one respect. According to Walkenbach (2005) the construction of spreadsheet formulas can be construed to be a type of programming." An appropriate development process for spreadsheet models provided by Grossman (2004) is very similar to the system development life cycle for traditional software. Robertson (2000) provides the typical first step in the software development process as being the realisation that all programs can be reduced to IPO.
- Inputs
- Processing (steps required to generate the desired outputs)
- Outputs

While spreadsheet models are sometimes criticised by traditional programmers for combining the processing and output components (and thus confusing matters), the IPO model is still a useful starting point.

Powell and Baker (2004) provide a fairly typical set of steps for engineering a spreadsheet model in the typical recommended layout. Included in the basic steps recommended by these authors is use of the concept of modularisation to separate data (inputs), decision variables, detailed calculations (processing) and outcome measures (output). They emphasise the importance of isolating input data or input parameters in a single location in the model (e.g. on one or more separate worksheets in a large model). Reusability of code is certainly one of the basic concepts of traditional programming that is partially achieved by the separation of data inputs from processing and outputs.

The fact that many spreadsheet models fail to adhere to one or both of these basic software design principles is almost certainly a major contributor to the incidence of spreadsheet errors and the lack of reusability of many models.

Substantiation for this the relevance of this paper is in essence provided by Powell and Baker (2004, page 97) with their explanation of why it is important to isolate input data or parameters.

> "A common source of errors in spreadsheets is the tendency to bury parameters in cell formulas and to replicate the same parameter in multiple cells. This makes identifying parameters difficult, because they are not immediately visible. It's also difficult to know whether all numerical values of a parameter have been changed each time an update is required. By contrast, the habit of using a single and separate location considerably streamlines the building and debugging of a spreadsheet."

**2.3 Academic Viewpoint of the Hard Coding Issue**

Based on the spreadsheet error classification method provided by Rajalingham (2005) it is not immediately apparent as to how hard-coding errors should be classified. Rajalingham borrows from Panko (1996) and defines quantitative errors as "numerical errors that lead to incorrect bottom-line values". It might therefore appear that hard coding errors should be defined as quantitative errors on the basis of this possible effect.

However it turns out Rajalingham classifies incidences of hard coding as qualitative errors on the basis that they do not necessarily result in bottom-line misstatement. This classification is logical in the sense that the type of hard coding of data values being considered in this research will never immediately result in a quantitative error.

Rajalingham (2005) does not seem overly concerned with the future effects of hard coding. He describes quantitative errors generally as to "degrade the quality of the model" and to make it "more prone to misinterpretation, and difficult to update / maintain". It is significant that the possibility of future bottom-line misstatement is not mentioned. Panko would appear to share Rajalingham's disregard for hard coded data input values. This statement is made in light of Panko's failure to even report the incidence of hard coding errors in either Panko and Halverson (1997) or Panko (2000). Panko's ignoring of hard coding errors is perhaps even more significant that Rajalingham.

Within Rajalingham's classification of accidental errors this research is concerned solely with structural errors. Accidental errors as a result of end users inputting erroneous data to the model (data input errors) are not considered in this study. Within his quantitative error classification Rajalingham identifies a single type of error that could be referred to as hard-coding. Within his modification errors category Rajalingham provides an example of a formula being overwritten with data. While this research recognises this error type as being significant its identification is straight forward and is achieved by the spreadsheet auditing packages that were examined (e.g. Spreadsheet Detective).

O'Beirne (2005) provides a comprehensive list of techniques for avoiding and identifying spreadsheet errors. An excellent description of how to use Excel's auditing tools is given but is lacking in any automated method for locating hard coding except for the extreme case of the hard coding of a value into a cell where a formula should exist.

Rajalingham, Panko and O'Beirne appear to share the view that hard coded input values are not of great issue. Powell and Baker are the exceptions with regards to their recognition of the importance of hard coding errors.

**2.4 Practitioner Viewpoint of Hard Coding**

Panko (2005) discusses the importance of manual inspection of spreadsheet cells as a means of accuracy verification. He does however conclude that while locating spreadsheet errors through manual formula inspection is a useful but tedious and time-consuming task . I would certainly confirm this observation particularly on the tedious aspect. Thankfully spreadsheet auditing software is available for automation of some tasks and streamlining of others. This study contends that the importance that commercial auditing software places on identifying and dealing with hard coding instances will be a rough proxy for the practitioner or business viewpoint.

There exist a fairly large number of commercially available spreadsheet auditing software packages. As part of the proposed PhD research the author intends to conduct a comprehensive analysis of the error detection capabilities of a wide selection of the available software. To date the capabilities of three spreadsheet auditing software have examined with regards to their capacity for detecting the hard coding of input values into formulas. Following is a summary of this preliminary (and far from complete) review.

Table 1  Ability to Identify Hard Coded Input Values

| Auditing Software Program | Ability to Identify Hard Coded Input Values |
|---|---|
| UltraSleuth Gold version 3.0 | No |
| Spreadsheet Detective version 2004 | Yes |
| EXChecker version 2.4 | Yes |

UltraSleuth Gold provides a number of useful features including a password remover which performed nicely on the author's first attempt at removing the password from a protected worksheet. The advanced search facility provided by this package provides a means for locating a specified string (including wildcards) inside formulas. However, testing of this feature revealed that while the 'specified string' can be a constant there is no way to identify formulas containing an unknown hard coded input value. UltraSleuth's search feature is therefore not useful for our purposes as the input value must be specified as the string being sought and in most cases the input value being sought will be unknown.

Spreadsheet Detective provides a comprehensive facility for detecting spreadsheet errors and additionally provides 'other checks and warnings' that highlights dubious constructs in the design of the spreadsheet model. This software package does not emphasise the potential quantitative errors caused by its identification of hard coded constants as it simply refers to them as being 'bad style'

The 'numeric constants' feature of compassoft's EXChecker v2.4 "provides a distinct list of all formulas that use a hard-coded numeric constant" (Compassoft Inc. 2005, p. 53). However the documentation provided for the analysis supplied by this feature is consistent with Panko (2000) and Rajalingham (2005) in that fails to highlight potential errors that may result. The user guide merely points out that the formulas identified by the 'numeric constants' feature may not be errors and that named values should be used.

The results of the academic and auditing software's position on hard coding of input data into formulas is summarised in table 2 following. This table illustrates quite clearly that while this practice is recognised as a design flaw it is not generally regarded to be a serious problem. The remaining portion of this study still to be conducted is to analyse a sample of spreadsheet models obtained from students, academics and practitioners. The AHCIM will be used to identify instances of hard coding and the errors caused by their incurrence.

**Table 2  Summary of Academic and Auditing Software Evaluations**

| Source | Evaluation of the Importance (Danger) of Formulas Containing Hard Code Input Values |
|---|---|
| **Researcher / Academic** | |
| Panko | Not important  - Not reported |
| Panko and Halverson | Not important - Not reported |
| Rajalingham | Not important - Qualitative error only |
| Powell and Baker | Important – Identified as a common source of error |
| O'Beirne | Not important – no method to detect |
| | |
| **Software Package** | |
| UltraSleuth Gold | Not important – No facility to identify |
| Spreadsheet Detective | Design issue only |
| Compassoft (EXChecker) | Design issue only |

## 3. AUTOMATED HARD CODING IDENTIFICATION METHOD (AHCIM)

The automated hard coding identification method (AHCIM) has been developed as a subroutine or procedure in Excel using VBA (Visual Basic for Applications) code. This procedure iterates through the cell collection of a specified range. If the cell has a formula, the formula is examined for operators (+, - / * etc). If an operator is found, the character immediately to the right is examined. If that character is numeric (and therefore not a cell address in the form "X9", "R9C9", "$X$9") that cell is flagged as potentially (probably ?) containing a constant that should be properly designed as a cell reference to the input parameter. AHCIM has also been programmed to detect instances of direct entry of numerical values into cells with the reasoning that such entries should not occur outside of the model's assumption or data area.

The current version of AHCIM does not provide an option for the user to exclude a region of the model to be excluded from the detection process. AHCIM recursively examines each worksheet to identify cells containing formulas. Each formula cell is then evaluated with regards to the existence of numbers within the formula. AHCIM relies on the premise that a properly constructed, general formula will never include data values but will consist of cell references to the spreadsheet model's data region. Formula cells containing data values will be identified by VBA programming code that searches each formula for numbers (integers) that are preceded by a mathematical operator (e.g. =, +, -, *, /). This search recognises that numbers (integers) for cell references (e.g. +B22) will exist in proper, general formulas but will never be immediately preceded by an operator.

Testing and debugging of AHCIM revealed several complications that may be introduced by the modeller's selection of formatting and security measures. The contents of cells can be hidden from user view in a variety of ways (e.g. hidden cells, rows, columns, font colours). Other security measures can be invoked through VBA code that hide formulas, cell contents and very hidden worksheets. Merged cells also caused a problem with the initial prototypes of AHCIM.

As AHCIM will identify formulas that contain any constants it will flag formula that contain 0 or 1 which may be properly required by the formula (e.g. 1 – tax rate %). Our logic for not having the procedure ignore these constants was the thinking that we it is better to identify a potential problem that can be dismissed than to mistakenly not identify a potential problem formula.

As searching every cell on a worksheet is a time consuming activity (256 columns X 8192 rows) the VBA *UsedRange* property for identifying the area of the worksheet that has been used by the developer. This property allows the search process to be limited to a cell region substantially less than 256 columns and 8192 rows.

## 4. DISCUSSION AND RESULTS FROM USE OF AHCIM

### 4.1 Analysis of Student Assignment Spreadsheet Models

Following is a screen shot of the analysis worksheet created by AHCIM for its evaluation of the workbook submitted by a management accounting student as their completed assignment of a comprehensive budgeting exercise. The summary information provided in the top four rows of the worksheet provides the name and location of the workbook examined together with an indication of the number of worksheets examined (11), the number of formulas examined (481), the instances of hard coding identified (88) and the number of numeric values entered directly to cells (50). While such numeric values are not "hard codings" they are identified by AHCIM to allow the user to verify that these entries are expected being either input or assumption data.

The screen shot below illustrates an illustrative sample of the 68 instances of "hard coding" identified (the rows for instances 7-28, 31-40 and 55-64 have been hidden as they are repetitive). Six of the identified hard codings are actually numerical values entered directly to cells (instances 1-2 and 65-68) as is apparent as the cell formula is the same as the cell value for these items. While instances 65-68 are acknowledged as being appropriate input data values (e.g. located on the model's Data worksheet) instances 1&2 identify errors by the student where numerical answers have been provided instead of cell referenced formula.

Instance 3 (cell D16 on worksheet C) reflects a true hard coding that would result in a future bottom-line error. The student has typed the number 200 into their formula instead of referencing to the cell containing this *minimum number of units in finished goods inventory* requirement on the assignment's data worksheet. Future changes in inventory policy may result in the student's model producing erroneous production outputs. Table 3 shows further hard codings of this 200 unit inventory minimum in instances 4 and 6.

The remainder of the hard coding instances have identified the student's use of the constants 0, 1 and 12 which require only cursory examination to determine their appropriateness.

**Table 3 Screen Shot of AHCIM Analysis of a Single Spreadsheet Model**

| No. | Worksheet | Cell | Cell Formula | Cell Value | #1 | #2 | #3 | #4 |
|---|---|---|---|---|---|---|---|---|
| 1 | ExecSummary | $A$14 | 9732311 | 9732311 | | | | |
| 2 | ExecSummary | $H$14 | 101707.5 | 101707.5 | | | | |
| 3 | C | $D$16 | =IF(C17*Data!$C$30>200, Data!$C$30*C1C17, 200) | 382 | 200 | | | |
| 4 | C | $D$17 | =IF(C6*(Data!H9+1)*Data!C30>200, C6*(Data!H9+1)*Data!C30, 200) | 247.536 | 1 | 200 | 1 | |
| 5 | Prod 2007 | $C$6 | ='C1C6*(1+Data!$H$9) | 618.84 | 1 | | | |
| 6 | Prod 2007 | $D$6 | =IF(C7*Data!$C$30>200, Data!$C$30*C7, 200) | 288.792 | 200 | | | |
| 29 | F | $J$7 | =I7*(1-Data!$C$24) | $11,271.38 | 1 | | | |
| 30 | F | $J$8 | =I8*(1-Data!$C$24) | $15,814.80 | 1 | | | |
| 41 | cp | $E$7 | =Data!$E$35/12 | $30,000.00 | 12 | | | |
| 53 | I | $G$7 | =IF(F7<0, F7*Data!$E$39/12, 0) | ($69.12) | 0 | 12 | | |
| 54 | I | $G$8 | =IF(F8<0, F8*Data!$E$39/12, 0) | ($97.71) | 0 | 12 | | |
| 65 | Data | $C$5 | 9550 | 9550 | | | | |
| 66 | Data | $C$7 | 175 | $175.00 | | | | |
| 67 | Data | $H$9 | 0.08 | 0.08 | | | | |
| 68 | Data | $C$10 | 0.06 | 0.06 | | | | |

Header info: Workbook Name: 305161814RowNumber7.xls; Workbook Location: H:\BudgetCase\Standardised; Wks hts: 11; F'mulas: 481; Hard Codes: 88; Num'c Values: 50.

The AHCIM procedure has been programmed to allow selection of multiple spreadsheet models for analysis at one time with an analysis worksheet in the form displayed in table 3 above being created for each model (the maximum number of models that can be selected at one time being limited by Excel's worksheet maximum of 256). In addition to the individual model analysis worksheets, a summary worksheet is also produced affording a view of the overall analysis performed as illustrated in table 4 below.

Table 4  Screen Shot of AHCIM Analysis Summary

| | Workbook Name | Workbook Location | # worksheets | # formulas | # hard codings | # numeric values |
|---|---|---|---|---|---|---|
| #1 | Budget Case-HongY.xls | P:\AtUni\ContEducation\BudgetCaseSubmissions | 12 | 468 | 91 | 43 |
| #2 | Budgeting assignment_Liu Liu.xls | P:\AtUni\ContEducation\BudgetCaseSubmissions | 12 | 496 | 60 | 45 |
| #3 | Jason324442.xls | P:\AtUni\ContEducation\BudgetCaseSubmissions | 13 | 821 | 47 | 42 |
| #4 | KejiaLi347280.xls | P:\AtUni\ContEducation\BudgetCaseSubmissions | 12 | 478 | 87 | 43 |
| #5 | KejiaLi347280revised.xls | P:\AtUni\ContEducation\BudgetCaseSubmissions | 12 | 479 | 78 | 43 |
| #6 | SUNyan338091.xls | P:\AtUni\ContEducation\BudgetCaseSubmissions | 12 | 513 | 66 | 42 |
| #7 | TinaJeng348737.xls | P:\AtUni\ContEducation\BudgetCaseSubmissions | 12 | 558 | 78 | 48 |
| #8 | TinaKwok348737.xls | P:\AtUni\ContEducation\BudgetCaseSubmissions | 12 | 558 | 78 | 48 |
| #9 | WeatherburnManagementaccountingassign.xls | P:\AtUni\ContEducation\BudgetCaseSubmissions | 12 | 857 | 128 | 43 |

The analysis summary provided in table 4 above was produced from the analysis of nine spreadsheet models submitted by management accounting students.  This summary provides an interesting overview especially when each of the models analysed (as in the case) should be providing similar outputs based on identical input data.

Table 4 provides analysis for a budgeting assignment where the template provided to each student contained 42 data input cells.  All of the analysed submissions except for student #3 and #6 have entered numerical values into cells that warrant investigation as the assignment template provided all required data.  The disparity in the number of formulas used to complete the assignment is interesting and worthy of further research with regards to output accuracy and usability as a function of model design.

**4.2  AHCIM Analysis of a Complex Practitioner Model**

Table 5 below provides a summary view of the AHCIM analysis of a practitioner spreadsheet model used to assist financial analysts in development of their own analytical models.  This highly complex model consists of 18 separate worksheets and over 50,000 formulas.  Examination of the summary analysis provided in table 5 would result in a questioning of the logic underlying some of the less obvious constants.  Not withstanding the response to such queries it would appear that this model is well designed with regards to its use of proper modelling techniques to avoid the hard coding of constants into formulas.  However investigation of the meaning of the 'magic numbers' (e.g. 236, 259, 279) is certainly warranted and could potentially reveal a serious problem.

Table 5  AHCIM Analysis Summary of Complex Practitioner Spreadsheet Model

Summary of AHCIM Analysis

| # worksheets | # formulas | # hard codings | # numeric values |
|---|---|---|---|
| 18 | 51183 | 14578 | 10461 |

| Constant Value | Number of Occurrences | Constant Value | Number of Occurrences |
|---|---|---|---|
| 0 | 4095 | 12 | 869 |
| 0.01 | 38 | 100 | 145 |
| 1 | 6278 | 236 | 39 |
| 2 | 937 | 237 | 40 |
| 3 | 114 | 259 | 61 |
| 4 | 174 | 279 | 55 |
| 5 | 20 | 1000 | 1438 |
| 6 | 242 | 1000000 | 11 |

## 5. SUMMARY AND FUTURE RESEARCH

This research begins with the proposition that errors caused by spreadsheet models containing formulas with hard coded input values are severely underestimated. It further hypothesises that the incidence and magnitude of these errors will increase with the age of the model. While this paper does not provide empirical evidence to support either of these propositions it has provided the groundwork for doing so with its development of the AHCIM procedure. The expectation is that this future research with AHCIM will provide strong empirical evidence to support both of these hypotheses. These future results will provide a solid foundation for reemphasising the importance of appropriate design techniques and structure when developing spreadsheet models.